\documentclass[10pt,twocolumn]{IEEEtran}

\IEEEoverridecommandlockouts

\usepackage{amsthm, amssymb}

\newtheoremstyle{slplain}
  {3pt}
  {3pt}
  {\slshape}
  {}
  {\bfseries}
  {.}%
  { }
  {}

\theoremstyle{slplain}

\usepackage{cite}
\usepackage{amsfonts}
\usepackage{multicol,multienum}
\usepackage{multirow}

\usepackage[ruled]{algorithm2e}

\ifCLASSINFOpdf
  \usepackage[pdftex]{graphicx}
  \graphicspath{{../pdf/}{../jpeg/}}
  \DeclareGraphicsExtensions{.pdf,.jpeg,.png}
\else
  \usepackage{float}
  \usepackage[dvips]{graphicx}
  \graphicspath{{../eps/}}
  \DeclareGraphicsExtensions{.eps}
\fi

\usepackage{epstopdf}

\usepackage[cmex10]{amsmath}
\usepackage{algorithmic}
\usepackage{array}
\usepackage{url}
\usepackage{caption}

\begin{document}

\title{Random Access Preamble Design and Detection for 3GPP Narrowband IoT Systems}

\author{
\IEEEauthorblockA{Xingqin Lin, Ansuman Adhikary, and Y.-P. Eric Wang}
\thanks{X. Lin, A. Adhikary, and Y.-P. E. Wang are with Ericsson Research,
San Jose, California, USA. (Email: \{xingqin.lin, ansuman.adhikary, eric.yp.wang\}@ericsson.com). 
}
}

\maketitle

\begin{abstract}
Narrowband internet of things (NB-IoT) is an emerging cellular technology that will provide improved coverage for massive number of low-throughput low-cost devices with low device power consumption in delay-tolerant applications. A new single tone signal with frequency hopping has been designed for NB-IoT physical random access channel (NPRACH). In this letter we describe this new NPRACH design and explain in detail the design rationale. We further propose possible receiver algorithms for NPRACH detection and time-of-arrival estimation. Simulation results on NPRACH performance including detection rate, false alarm rate, and time-of-arrival estimation accuracy are presented to shed light on the overall potential of NB-IoT systems. 
\end{abstract}


\IEEEpeerreviewmaketitle

\section{Introduction}

In this letter, we study random access preamble design and detection for narrowband internet of things (NB-IoT) systems. Internet of Things (IoT) is a vision for the future world where everything that can benefit from a connection will be connected. Cellular technologies are being developed or evolved to play an indispensable role in the IoT world \cite{andreev2015understanding}. NB-IoT, being developed in the 3rd generation partnership project (3GPP), promises to provide improved coverage for massive number of low-throughput low-cost devices with low device power consumption in delay-tolerant applications \cite{3gppIotRWI}.

The system bandwidth of NB-IoT is 180 kHz for both downlink and uplink. 
The downlink transmission is based on conventional orthogonal frequency-division multiple access (OFDMA) with 15 kHz subcarrier spacing. 
Both single tone and multi-tone transmissions are supported in the uplink. The subcarrier spacing for the single tone transmission can be either $15$ kHz or $3.75$ kHz. The multi-tone transmission is based on single-carrier frequency division multiple access (SC-FDMA) with 15 kHz uplink subcarrier spacing \cite{3gppIotRWI}. In this letter, we focus on the NB-IoT physical random access channel (PRACH), known as NPRACH. NPRACH refers to the time-frequency resource on which random access preambles are transmitted. Transmitting a random access preamble is the first step of random access procedure that enables a user equipment (UE) to establish a connection with the network. Acquiring uplink timing is another main objective of random access in OFDMA systems. The acquired uplink timing is used to command the UE to perform timing advance to achieve uplink synchronization in OFDMA systems \cite{dahlman20134g}.

In LTE, a set of random access preambles based on Zadoff-Chu (ZC) sequences is configured within a cell \cite{3gpp36211}. Though ZC sequences are constant-envelope, the PRACH signal after upsampling and filtering has a peak-to-average power ratio (PAPR) in the range of 2 to 7 decibels \cite{lotter1995constant}. The higher the PAPR, the more the required power amplifier backoff. Power amplifier backoff also gives rise to degraded power amplifier efficiency, and thus has a negative impact on device battery life time.

A new single tone PRACH signal with frequency hopping has been designed for NPRACH in NB-IoT \cite{3gppIotCR}. This new PRACH signal (described in Section \ref{sec:design}) has extremely low PAPR and thus significantly reduces the need for power amplifier backoff. 
NPRACH also aims to improve the coverage of random access by more than 20 dB compared to GSM/GPRS. 
It helps fulfill the important performance objectives of IoT such as long battery lifetime and extended coverage. Existing works (see e.g., \cite{laya2014random, jang2014spatial, oh2015joint} and references therein) on random access for machine-to-machine (M2M) or IoT applications are centered on traditional LTE-type ZC based PRACH signal. In this letter, we describe this new NPRACH design, explain the design rationale, and discuss possible receiver algorithms for NPRACH detection.

\section{Random Access Preamble Design}
\label{sec:design}


Consider an OFDMA system with $W$ Hz bandwidth and $B$ Hz subcarrier spacing. The fast Fourier transform (FFT) size, denoted as $N$, is usually chosen as a power of $2$ larger than the number $W/B$ of subcarriers. Accordingly, the baseband sampling rate can be conveniently chosen as $NB$ Hz. For NPRACH in NB-IoT with $W=180$ kHz and $B=3.75$ kHz, there are $48$ subcarriers available and we may choose $N=64$. 

\begin{figure}
\centering
\includegraphics[width=8cm]{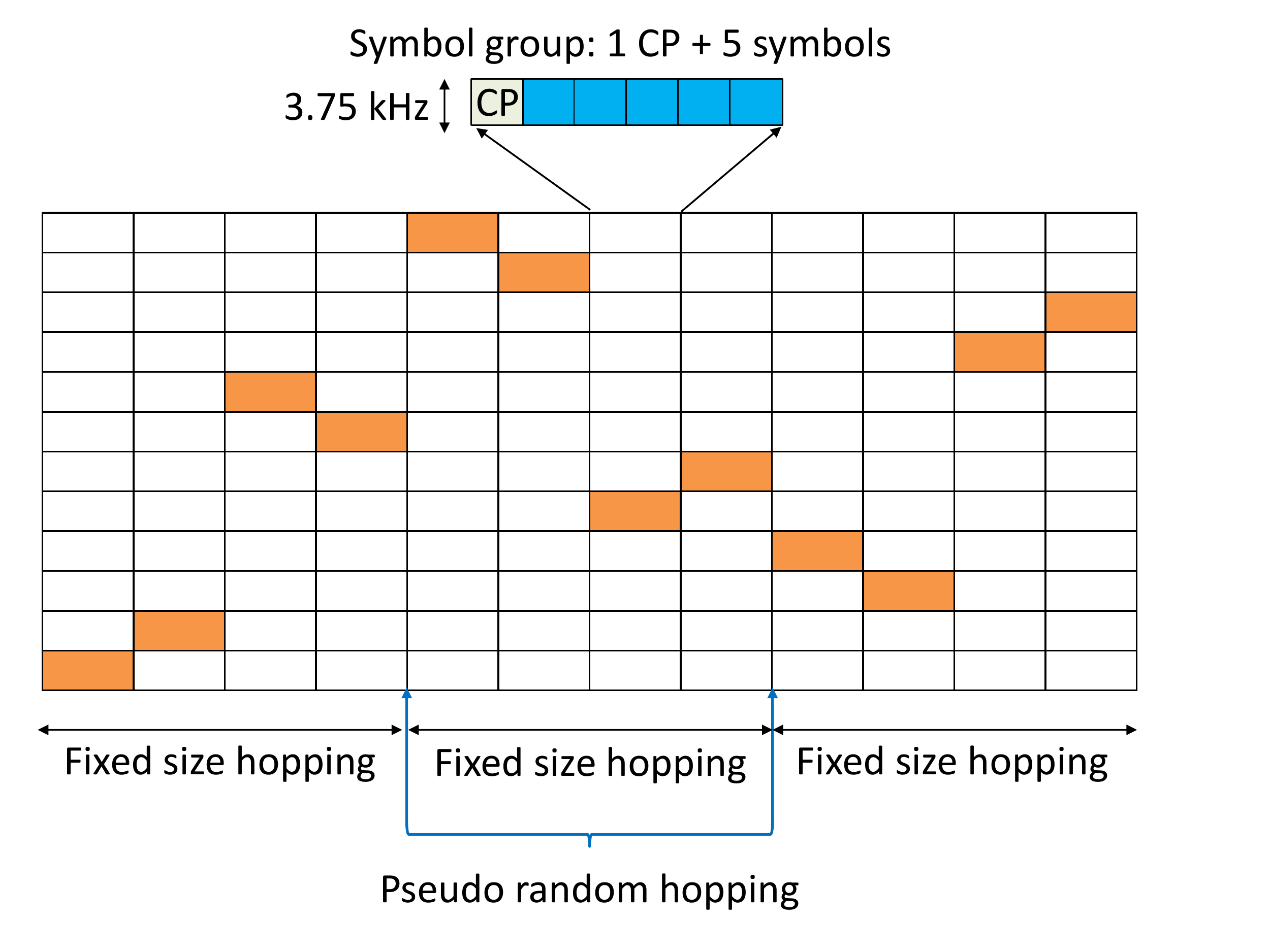}
\caption{Illustration of 3GPP NB-IoT physical random access channel design.}
\label{fig:1}
\end{figure}

The random access preamble design is based on single subcarrier transmission with frequency hopping within a configured NPRACH band within the OFDM resource grid, as illustrated in Figure \ref{fig:1}.

\subsection{Symbol Group}

A classical OFDM symbol structure consists of a cyclic prefix (CP) portion and a data symbol. In order to maintain the orthogonality of the random access transmissions on different subcarriers, the CP has to be long enough to accommodate the timing uncertainty in the cell that can be as large as the maximum round-trip delay (plus additional channel delay spread and downlink synchronization errors). Cellular IoT/M2M systems normally target at large cell deployment. For example, 35 km cell size should be supported for NB-IoT \cite{3gppIotSI}, requiring a CP at least as long as $233.3$ us, which is comparable to the duration of a data symbol with $B=3.75$ kHz subcarrier spacing. To reduce the relative CP overhead, we can repeat each $N$-sample OFDM symbol $\xi$ times, and then add a single CP of $N_{cp}$ samples. As a result, the length of the composite symbol is $N_{cp} + \xi N$ samples. The larger the number $\xi$ of repetitions, the smaller the CP overhead. The value of $\xi$ however should be kept small enough such that the effect of channel variation is negligible within $N_{cp} + \xi N$ samples. Otherwise, undesirable inter-carrier interference (ICI) would arise.

To avoid confusion with the classical notion of OFDM symbol, we refer to the composite symbol composed of $N_{cp} + \xi N$ samples as a \textit{symbol group}. Note that, when $\xi=1$, the symbol group structure reduces to the classical OFDM symbol structure. A random access preamble consists of $L$ symbol groups and uses one subcarrier at every symbol group for transmission. The length $L$ of the preamble is determined by the target operating signal-to-noise ratio (SNR) or equivalently, the target maximum coupling loss (MCL) in the networks. Note that NB-IoT supports different coverage classes up to $164$ dB MCL. The networks can configure different values of $L$ to cater for different coverage requirements.

In NPRACH, two CP lengths are specified: $266.7$ us and $66.7$ us. One example usage of the CP lengths may go as follows. The long CP of $266.7$ us is used for large cells with radii in the range $8-35$ km, and the short CP of $66.7$ us is used for cells with radii smaller than $8$ km. The number $\xi$ of symbols in a symbol group is specified to be $5$ \cite{3gppIotCR}.

\subsection{Hopping Pattern}

A prominent feature of NPRACH is the hopping pattern design, which is illustrated in Figure \ref{fig:1}. We can see that the hopping consists of both inner layer fixed size hopping and outer layer pseudo-random hopping. Outer layer pseudo-random hopping is applied between groups of 4 symbol groups. Inner layer fixed size hopping is applied within every 4 symbol groups.
\begin{itemize}
\item First level single-subcarrier hopping is used between the first and the second and between the third and the fourth symbol groups. Further, the two single-subcarrier hoppings are mirrored, i.e., if the first hopping is ``UP'', the second hopping is ``DOWN'', and vice versa.  
\item Second level 6-subcarrier hopping is used between the second and the third symbol groups.
\end{itemize}


The rationale behind the hopping pattern of NPRACH will become clear after we go through preliminary analysis in Section \ref{subsec:preAna}. For the sake of description, we consider a generic hopping pattern. Specifically, denote the subcarrier index used by symbol group $m$ as $\Omega(m), m=0,...,L-1$, where $\Omega(\cdot)$ is a generic mapping from symbol group index to the subcarrier index. The specific form of $\Omega(m)$ is determined by the adopted hopping pattern, e.g. the multi-level hopping adopted in NPRACH.

\section{Design Rationale Explained}
\label{sec:explained}

\subsection{Preliminary Analysis}
\label{subsec:preAna}

Without loss of generality we assume $\xi=1$ in the following analysis. We will bring back the role of $\xi$ later. Denote by $\{u[m]\}_{m=0}^{L-1}$ the sequence of a random access preamble. The baseband equivalent digital-domain signal for the random access preamble transmission can be written as
\begin{align}
&s[n;m]  =  \frac{\sqrt{E}}{N} \sum_{k} S [k; m] e^{ j 2 \pi \frac{k}{N} n } , n= -N_{cp},..., N-1, \notag 
\label{eq:2}
\end{align}
where $s[n;m]$ denotes the $n$-th time-domain sample of the $m$-th symbol group, $E$ denotes the transmit energy per sample, and $S[k;m]$ denotes the symbol on the $k$-th subcarrier during the $m$-th symbol group.

For the narrow random access hopping range that is not larger than $45$ kHz in a NB-IoT system, the channel can be sufficiently modeled as a one-tap channel. Denoting by $h[n;m]$ the channel coefficient at the $n$-th time-domain sample of the $m$-th symbol group,
\begin{align}
h[n;m]=a[m]\delta [n] ,
\end{align}
where $a[m]$ is the channel gain at the $m$-th symbol group. The implicit assumption in the channel model is that the channel is invariant within one symbol group. This is a common assumption for OFDM transmissions. 

Denote by $D$ the round-trip delay, i.e., time-of-arrival (ToA), to be estimated by the base station (BS) for timing advance command. We assume that by design the CP is long enough to cover the maximum round-trip delay, and thus $D \in [0, N_{cp}-1 ]$. Accordingly, the $n$-th sample of the $m$-th symbol group at the receiver is given by
\begin{align}
y[n;m] 
= & a[m] \frac{\sqrt{E}}{N} e^{ j 2 \pi \Delta f ( n-D + m(N_{cp}+ N) ) }   \notag \\
&\times \sum_{k} S [k; m] e^{ j 2 \pi \frac{k}{N} (n-D) } + v[n;m] , 
\label{eq:3}
\end{align}
where $\Delta f$ is the residual carrier frequency offset (CFO) normalized by the sampling rate $NB$, and $v[n;m]$ is a complex additive white Gaussian noise (AWGN) with zero mean and variance $N_0$, i.e., $v[n;m] \sim \mathcal{CN}(0,N_0)$. The residual CFO may be due to imperfect carrier frequency estimation in the downlink cell search.

For each symbol group with $\xi=1$, the receiver discards the first $N_{cp}$ samples and performs a FFT on the remaining $N$ samples. 
At symbol group $m$, if $\ell = \Omega (m)$, it can be shown that the received symbol on the $l$-th subcarrier at symbol group $m$ is given by
\begin{align}
Y[\ell; m] 
=& B(\Delta f, D) a[m] u[m] e^{ j 2 \pi \Delta f  m(N_{cp}+ N)  } e^{ -j 2 \pi \frac{\pi(m)}{N} D   } \notag \\
 &+ V[\ell;m],
\label{eq:4}
\end{align}
where
\begin{align}
B(\Delta f, D)  = \sqrt{E} \frac{\sin(N\pi \Delta f)}{N \sin(\pi \Delta f)}
 e^{ j 2 \pi \Delta f ( \frac{N-1}{2}  -D )  } .
\end{align}



It is straightforward to extend (\ref{eq:4}) with $\xi=1$ to the more general case with $\xi\geq 1$. The difference is that the BS receives $\xi$ symbols for every symbol group. With a slight abuse of notation, the $i$-th received symbol in symbol group $m$ is given by 
\begin{align}
\tilde{y}[i; m] 
=& B(\Delta f, D) a[m] u[m] e^{ j 2 \pi \Delta f  \left( m(N_{cp}+ \xi N) + i N) \right)  } \notag \\
&\times
 e^{ -j 2 \pi \frac{\Omega(m)}{N} D   }   
 + \tilde{v}[i;m]  .
\label{eq:6}
\end{align}

\subsection{Design Rationale of NPRACH Frequency Hopping}

Now we are in a position to discuss the rationale behind NPRACH hopping pattern, which is illustrated in Figure \ref{fig:1}. From Eq. (\ref{eq:6}), it can be seen that the effect of ToA $D$ is captured in the term $e^{ -j 2 \pi \frac{\Omega(m)}{N} D   }$. With hopping (i.e., varying $\Omega(m)$) the phases vary across symbol groups. Therefore, the BS can estimate ToA by processing the phases of the received symbols. Some remarks on $e^{ -j 2 \pi \frac{\Omega(m)}{N} D   }$ are in order.

\begin{itemize}
\item Since the phase difference of two adjacent received symbol groups $m$ and $m+1$ due to hopping is proportional to the hopping distance $|\Omega(m+1)-\Omega(m)|$, choosing a larger hopping distance provides a finer resolution for ToA estimate and thus better estimation accuracy.
\item The phase difference of two adjacent received symbol groups $m$ and $m+1$ due to hopping is prone to $2\pi$ phase ambiguity, which may cause ambiguity in the ToA estimation. To avoid the $2\pi$ phase ambiguity, a larger hopping step $|\Omega(m+1)-\Omega(m)|$ reduces the ToA estimation range, which in turn translates into smaller cell sizes that can be supported.
\end{itemize}

From the above remarks, it seemed there was a design tradeoff between ToA estimation range and accuracy when choosing the frequency hopping steps. This issue can be resolved with multi-level frequency hopping patterns that include both small and large step hopping. In particular, the design rationale of NPRACH hopping is as follows. 
\begin{itemize}
\item Singe-subcarrier hopping ensures large enough ToA estimation range for NB-IoT target cell sizes (up to 35 km).
\item Six-subcarrier and pseudo random hopping improve ToA estimation accuracy.
\item Pseudo random hopping brings in additional system level benefits including reduced inter- and intra-cell interference, reduced vain responses to preamble transmissions in neighboring cells, etc.
\end{itemize}
With the multi-level frequency hopping, NPRACH design supports large ToA estimation range (which is needed for large cell size) while enabling acceptable ToA estimation accuracy at the BS.

\section{Receiver Algorithms}
\label{sec:algorithms}


Based on the results and discussions in Section \ref{sec:explained}, here we discuss how the BS can detect the random access preamble and estimate the ToA. We start with ToA estimation and then utilize the corresponding ToA estimation statistic to detect the presence of the preamble. 


Generally speaking, the task of ToA estimation is to estimate $D$ from the $\xi L$ observations $\{\tilde{y}[i; m]\}$ in the presence of additive noise and $L+1$ additional unknown parameters: $\{a[m]\}_{m=0}^{L-1}$ and $\Delta f$. In the initial random access, the BS normally does not have \textit{a priori} knowledge about the channel. Therefore, a joint estimator of all the parameters including $D$, $\{a[m]\}_{m=0}^{L-1}$, and $\Delta f$ would lead to too high complexity to be of practical use

To overcome this difficulty, we make the simplifying approximation of a block fading model: $\{a[m]\}_{m=0}^{L-1}$ do not change in a block of $Q$ symbol groups but change independently over the blocks. We choose $Q$ such that the number $L/Q$ of blocks is an integer. Mathematically, denoting by $\tilde{a}_g$ the nominal channel coefficient for block $g$, we have
\begin{align}
a[m] \equiv \tilde{a}_g, \quad \forall m= gQ,..., (g+1)Q -1,
\end{align}
and $\{\tilde{a}_g\}_{g=0}^{L/Q-1}$ are independent. Note that this is not a requirement on the system but an assumption in the algorithm used by the BS. 

With the block fading assumption, the ToA and residual CFO can be jointly estimated as follows.
\begin{align}
( D^\star, \Delta f^\star ) & = \arg \max_{D, \Delta f} J (D,\Delta f )  \notag \\
&= \arg \max_{D, \Delta f} \sum_{g=0}^{L/Q-1} \bigg | J_g (D,\Delta f )\bigg |^2,
\label{eq:5}
\end{align}
where
\begin{align}
J_g (D,\Delta f ) =& \sum_{m=gQ}^{(g+1)Q-1} \sum_{i=0}^{\xi-1}   \tilde{y}[i; m]
u^*[m] \notag \\ &\times
 e^{- j 2 \pi \Delta f  \left( m(N_{cp}+ \xi N) + i N) \right)  } e^{ j 2 \pi \frac{\Omega(m)}{N} D   }.
\end{align}

The above rule (\ref{eq:5}) of joint ToA and residual CFO estimation is intuitive. The estimate $(D^\star, \Delta f^\star)$ is the one that yields the maximum correlation of the transmitted preamble symbols and the received symbols whose phase shifts due to ToA and residual CFO are corrected by the estimate. Note that the estimation rule (\ref{eq:5}) takes the form of two-dimensional discrete-time Fourier transform (DTFT). As a result, the search for $(D^\star, \Delta f^\star)$ can be efficiently carried out in the frequency domain by utilizing an FFT. 


Assuming a joint estimate $(D^\star, \Delta f^\star)$ has been obtained, we next discuss how to determine the presence of the preamble. In LTE, the BS correlates the received signal with a hypothesis of Zadoff-Chu sequence based random access preamble \cite{dahlman20134g}. If the correlation result exceeds some predetermined threshold, the BS declares the presence of the preamble; otherwise, the BS declares that the preamble is not present. Similar approach can be used for the detection of NPRACH in NB-IoT. The natural choice of the statistic used to compare against the detection threshold is $J (D^\star,\Delta f^\star )$. It is well known that two error events may arise with this threshold-based preamble detection.  
\begin{itemize}
\item \textit{Misdetection}: The random access preamble is present, but the statistic $J (D^\star,\Delta f^\star )$ does not exceed the detection threshold.
\item \textit{False alarm}: The random access preamble is absent, but the statistic $J (D^\star,\Delta f^\star )$ exceeds the detection threshold.
\end{itemize}
Clearly, there exists a trade-off in setting the detection threshold. Increasing the detection threshold lowers the false alarm rate at the cost of increased likelihood of misdetection. For random access preamble detection in cellular systems, the detection threshold is usually chosen such that the false alarm rate is below some target. 

%

\section{Simulation Results}

In this section, we provide some simulation results to evaluate the NPRACH design. The simulation assumptions used are based on the ones outlined in \cite{3gppIotSI}, and are summarized in Table \ref{tab:sys:para}. We consider three coverage classes and configure $8$, $32$, and $128$ symbol groups for the three coverage classes, respectively. The target operating SNRs for the three coverage classes are $14.25$ dB, $4.25$ dB, and $-5.75$ dB, respectively. Note that according the evaluation setup agreed in \cite{3gppIotSI}, these three operating SNRs correspond to $144$ dB MCL, $154$ dB MCL, and $164$ dB MCL, respectively. While $144$ dB MCL is the coverage level reachable by GSM/GPRS systems, an additional $20$ dB coverage extension (i.e., $164$ dB MCL) is targeted in NB-IoT systems.

\begin{table}
\centering
\caption{Simulation Parameters}
\begin{tabular}{|l||r|} \hline
CP length & 266.7 us \\ \hline
Subcarrier spacing & 3.75 kHz \\ \hline
Symbol group & 1 CP and 5 symbols \\ \hline
NPRACH band & 12 subcarriers \\ \hline 
Channel model  & Typical urban \\ \hline 
Doppler spread  & $1$ Hz \\ \hline 
Antenna configuration  & 1 Tx; 2 Rx \\ \hline 
Timing uncertainty  & Randomly chosen between $0$ and CP \\ \hline 
Frequency error  & Uniformly drawn from $\{-50, 50\}$ Hz \\ \hline 
Frequency drift  & Uniformly drawn from $\{-22.5, 22.5\}$ Hz/s \\ \hline 
No. of iterations  & 10,000 for detection; 100,000 for false alarm \\ \hline 
\end{tabular}
\label{tab:sys:para}
\end{table}

\begin{figure}
\centering
\includegraphics[width=8cm]{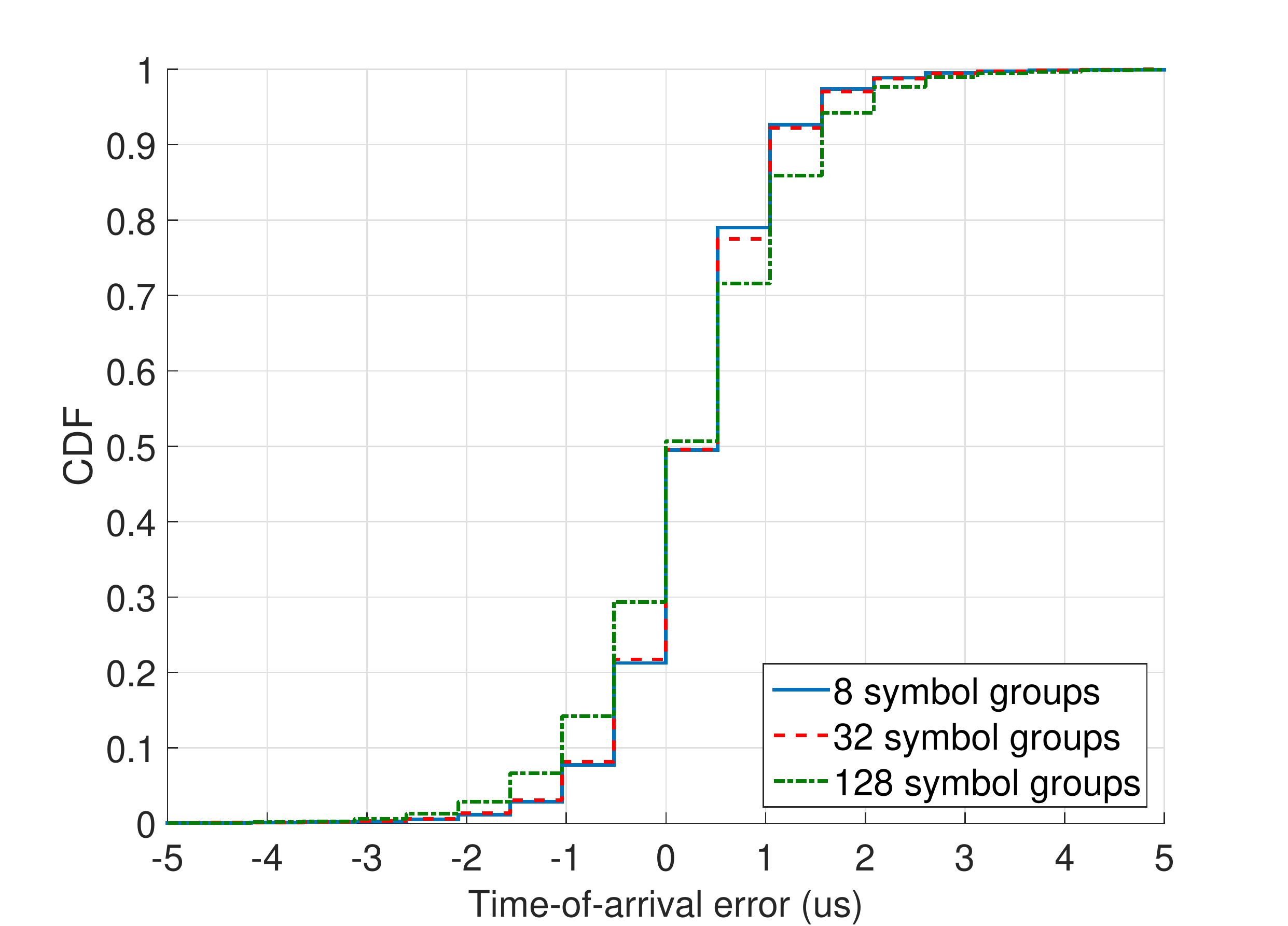}
\caption{Distributions of ToA estimation errors}
\label{fig:2}
\end{figure}

Table \ref{tab:sys:para} summarizes the misdetection and false alarm probabilities of the NPRACH design under the configured three coverage classes. It can be seen that the detection probabilities exceed $99\%$, while the false alarm probabilities are well below $0.1\%$. Figure \ref{fig:2} shows the distributions of ToA estimation errors for the three configured coverage classes. It can be seen that the ToA errors are within [-3, 3] us with very high confidence level for all three cases. These results demonstrate that NPRACH can fulfill the extended coverage requirement and help to achieve acceptable uplink synchronization accuracy.

\begin{table}
\centering
\caption{Misdetection and false alarm probabilities}
\begin{tabular}{|l||c|c|c|} \hline
 & Coverage 1 & Coverage 2 & Coverage 3 \\ \hline 
Target SNR  & 14.25 dB & 4.25 dB  & -5.75 dB \\ \hline 
No. of symbol groups   & 8 & 32 & 128 \\ \hline 
Misdetection   & 29/10,000 & 24/10,000 & 84/10,000 \\ \hline 
False alarm   & 0/100,000 & 0/100,000 & 13/100,000 \\ \hline 
\end{tabular}
\label{tab:detection}
\end{table}

\section{Conclusion}

In this letter, we introduced the new single tone frequency hopping random access signal used by NPRACH in NB-IoT systems. We explained in detail the design rationale and proposed some possible receiver algorithms for NPRACH detection and time-of-arrival estimation. We have also presented simulation results to shed light on NPRACH performance. Future work may consider developing more efficient and/or advanced receiver algorithms for NPRACH. 	One can also consider extending this work to study system level metrics such as preamble collision rate, random access capacity, and overall NB-IoT capacity that takes into account both random access channel and data channel.

\bibliographystyle{IEEEtran}
\bibliography{IEEEabrv,Reference}

\end{document}